\documentclass{cjpsuppl}
 \usepackage{graphics} 

\begin{document}
 \slacs{.6mm}
 \title{Spin dependent elastic antiproton interactions}
 \authori{D.~S.~O'Brien$^*$\footnotetext{$^*$ email: donie@maths.tcd.ie}{\ } and{\ } N.~H.~Buttimore}
 \addressi{School of Mathematics, Trinity College, Dublin, Ireland}
 \authorii{}
 \addressii{}
 \authoriii{}   \addressiii{}
 \authoriv{}    \addressiv{}
 \authorv{}     \addressv{}
 \authorvi{}    \addressvi{}
 \headtitle{Spin dependent elastic antiproton interactions}
 \headauthor{D.~S.~O'Brien, N.~H.~Buttimore}
 \lastevenhead{D.~S.~O'Brien, N.~H.~Buttimore: Spin dependent elastic antiproton interactions}
 \refnum{}
 \daterec{} 
 \suppl{?}  \year{2006} \setcounter{page}{1}
 \maketitle

 \begin{abstract}
Relativistic formulae for spin averaged and spin dependent one photon exchange differential cross sections are developed for spin half fermion fermion elastic scattering.  Spin transfer cross sections are important for the Polarized Antiproton eXperiments (PAX) project at GSI Darmstadt.  In particular, cross sections for polarization transfer in antiproton electron $\bar{p} \,  e \!\!\uparrow \;\longrightarrow \bar{p} \!\uparrow \! e$ and antiproton proton $\bar{p} \,  p \!\!\uparrow \;\longrightarrow \bar{p} \!\uparrow \!p\,$ elastic collisions are presented.

 \end{abstract}

\section{Introduction} 
\noindent
Antiprotons were discovered fifty years ago.
A proposal to accelerate polarized antiprotons formed
by spin filtering using a polarized target \cite{Rathmann}
has been suggested for the
High Energy Storage Ring (HESR) at the
Facility of Antiproton and Ion Research (FAIR).
Such a programme would allow the evaluation of
previously unknown single and double spin observables \cite{Barone:2005pu},
opening a new window to antiparticle studies of a unique kind.
The transversity distribution
of the valence quarks in the proton
together with the moduli and the relative phase
of the time-like electric and magnetic form factors
$G_E$ and $G_M$ of the proton would be improtant candidates for early measurement with the unprecedented antiproton facility.
In polarized and unpolarized antiproton proton elastic scattering,
open questions such as the contribution from the odd charge-symmetry
Landshoff-mechanism at large $|t|$ and spin-effects in the extraction
of the forward scattering amplitude at low $|t|$ can be addressed.

We investigate the results of elastic scattering of antiprotons off electrons and protons, due to single photon exchange, first unpolarized in sections 2 and 3.  Fully relativistic polarization transfer differential cross sections are derived for elastic antiproton electron and antiproton proton Coulomb scattering in section 4.  In section 5 our results are summarized.

Emphasis is put on the spin transfer differential cross section for elastic antiproton proton scattering, as it has been suggested in a recent paper that electrons are not effective in transferring polarization to antiprotons in the kinematic region of the PAX programme \cite{Milstein:2005bx}.


\pagebreak

\section{Cross sections}
\noindent
The differential cross section is related to the helicity amplitudes $\mathcal{M}(\Lambda'\lambda';\Lambda\,\lambda)$
 by
\begin{equation}
s \,\frac{d\sigma}{d\Omega} =
\, \frac{1}{(8\pi)^2}\sum_{\lambda\lambda'\Lambda\Lambda'} \frac{1}{(2\lambda +1)(2\Lambda +1)}\,|\mathcal{M}(\Lambda'\lambda';\Lambda\,\lambda)|^2 
\end{equation}
where $\lambda,\Lambda$ and $\lambda',\Lambda'$ are the helicities of the initial and final particles respectively, and the $s$ and $t$ are Mandelstam variables  \cite{Mandelstam}.  The spin half current of a point particle, such as the electron current is
\begin{equation}
 j^\mu = e \,\bar u(k',\lambda') \, \gamma^\mu \, u(k,\lambda)
\end{equation}
where $k,P$ and $k',P'$ are the momenta of the initial and final particles respectively.
\\
Define electromagnetic form factors $F_1(q^2)$ and $F_2(q^2)$, with normalization $F_1(0)=1$ and $F_2(0)=\mu - 1$, the anomalous magnetic moment, where the momentum transfer to the photon $q= k' - k$  = $P - P'$ and $q^2 = t$.  Using $G_M = F_1 + F_2$  and Gordon decomposition, the proton current is
\begin{equation}
 J_\mu = e \,\bar u(P',\Lambda') \left( G_M \, \gamma_\mu 
- F_2\,\, \frac{P_\mu + P'_\mu}{2M} \,\right)u(P,\Lambda) 
\end{equation}
where $M$ is the mass of the proton.

\section{Spin averaged cases} 

 \subsection*{Structureless}
A standard calculation gives the differential cross section for one photon exchange, of two non-identical spin 1/2 point particles to be 
\begin{equation}
s\, \frac{d\sigma}{d\Omega} = \frac{\alpha^2}{2\,t^2}\left[\,2\left(s-m^2 
- M^2\,\right)^2  + 2\,s\,t +t^2\,\right]
\end{equation}
where $m$ and $M$ are the masses of the particles, and $\alpha = e^2/4\pi$.

 \subsection*{One particle structured}

Considering one particle to have structure determined by the electromagnetic form factors described in section 2, we obtain

\begin{eqnarray}
s\, \frac{d\sigma}{d\Omega}
& = 
& \frac{\alpha^2}{2\,t^2}\left\{G_M^{\,2}\left[\,2\left(s-m^2 - M^2\,\right)^2  + 2\,s\,t +t^2\, \right]
\phantom{\frac{t}{4M^2}}
\right.\nonumber \\& 
&  \left. - \,2\,F_2\left[F_2\left(1 + \frac{t}{4M^2}\right) + 2\,F_1\right]\left[ \left(s-m^2-M^2\,\right)^2 + t\left(\,s -m^2\,\right) \right]\right\}
\end{eqnarray}
a result that equals the previous result in the structureless limit $F_1\rightarrow 1$, $F_2 \rightarrow 0$ and hence $G_M \rightarrow 1$.  In the $m \rightarrow 0$ limit this is the Rosenbluth formula \cite{Rosenbluth}. 

 \subsection*{Both particles structured}
Defining the electromagnetic form factors $f_1(q^2)$ and $f_2(q^2)$ for the second particle and using $g_M = f_1 + f_2$ we obtain the spin averaged differential cross section due to one photon exchange
\begin{eqnarray}
s\,\frac{d\sigma}{d\Omega}
& =
& \frac{\alpha^2}{2\,t^2}\left\{
\,g_M^{\,2}G_M^{\,2}\left[\,2\left(s-m^2-M^2\,\right)^2
+
2\,s\,t + t^2\right]
\phantom{\frac{f_2F_2}{2}}
\right. \nonumber
\\
& & \left. -\,2\,g_M^{\,2}F_2\left[F_2\left(1 + \frac{t}{4M^2}\right) + 2\,F_1\right]\left[ \left(s-m^2-M^2\,\right)^2 + t\left(\,s -m^2\right)\right]\right. \nonumber \\& & \left. -\,2\,G_M^{\,2}f_2\left[f_2\left(1 + \frac{t}{4m^2}\right) + 2\,f_1\right]\left[\left(s-m^2-M^2\,\right)^2 +t\left(\,s - M^2\right)\right] \right.\\& &\left.
+\,\frac{f_2F_2}{2}\left[f_2\left(1
+ \frac{t}{4m^2}\right) + 2\,f_1\right]\left[F_2\left(1 + \frac{t}{4M^2}\right) + 2\,F_1\right](s-u)^2 \right\} \,.\nonumber 
\end{eqnarray}
Again this result equals the previous result in the one particle structured limit $f_1 \rightarrow 1$, $f_2 \rightarrow 0$ and hence $g_M \rightarrow 1$.  Here $u$ is the third Mandelstam variable.

 \subsection*{Antiproton proton case}
\noindent
For antiproton proton collisions, the electromagnetic form factors and masses of the proton and antiproton are the same, i.e. $f_1 = F_1$ and $f_2 = F_2$ so $g_M = G_M$; and $m = M$.  Here we neglect the $s$-channel one photon contribution in favour of the $t$-channel term which dominates in the low momentum transfer (small $t$) region of interest, and also dominates at high energies.  This gives
\begin{eqnarray}
s\,\frac{d\sigma}{d\Omega}
& = 
& \frac{\alpha^2}{2\,t^2}\left\{\,G_M^{\,4}\left[\,2\left(s-2M^2\,\right)^2 + 2\,s\,t + t^2\right]
\phantom{\left[F_2\left(1 + \frac{t}{4M^2}\right) + 2\,F_1\right]^2}
\right. \nonumber\\& 
& \left. -\,4\,G_M^{\,2}F_2\left[F_2\left(1 + \frac{t}{4M^2}\right) + 2\,F_1\right]\left[ \left(s-2M^2\,\right)^2 + t\left(\,s -M^2\right)\right]\right. \\& 
&   \left. \qquad \qquad \qquad \qquad \qquad+\,\frac{F_2^{\,2}}{2}\left[F_2\left(1 + \frac{t}{4M^2}\right) + 2\,F_1\right]^2(s-u)^2 \right\} \nonumber
\end{eqnarray}
in agreement with an expression formed from
known fermion fermion helicity amplitudes
\cite{Buttimore:1978ry},
and with a cross section formula for proton proton
scattering
\cite{Block:1996jd}.
In this case  $s+t+u=4M^2$ and thus $s-u = 2\,s+t-4M^2$.\\
\\
This result is important in the momentum transfer region $|t|<|t_c|$ for antiproton proton collisions with total cross section $\sigma_{\mathrm{tot}}$, defined by

\begin{equation}
t_c \,=\, - \,\frac{8\, \pi \, \alpha}{\beta_\mathrm{lab}\,\sigma_\mathrm{tot}} 
 \approx 
- \,0.001 \,\mbox{(GeV/}c)^2
\end{equation}
where the electromagnetic interaction dominates the hadronic interaction.  Here the laboratory velocity is $\beta_\mathrm{lab} \,=\,\sqrt{s(s-4M^2)}/(s-2M^2)$.

\section{Spin dependent cases} 
\noindent
Suppose the initial target electron (or proton) to have a spin four vector $S_\mu$ and the final scattered antiproton to have a spin four vector $S'_\mu$.  
\\
\\
We are most interested in the polarization transfer $K_{j00i}$, i.e. 
$
 \bar{p} \,  p \!\!\uparrow \;\longrightarrow \bar{p} \!\!\uparrow \!p\,
$.
First
\[		
	\bar p(P) + e(k,S)  \longrightarrow  \bar p(P',S') + e(k')
\] is investigated and then
\[		
	\bar p(P) + p(k,S)  \longrightarrow  \bar p(P',S') + p(k')\,.
\]
Consider now the spin dependent terms, and use the notation $ A \cdot B = A_\mu B^\mu$.

 \subsection*{Structureless}
\noindent
The cross section for polarization transfer from initial electron to final antiproton (assumed structureless here) involves the spin observable $K_{j00i}$, where $i$ and $j$ refer to the polarization directions as in reference \cite{Lehar:1978}.  We find the one photon exchange contribution to be

\begin{equation}
 s\,\frac{d\sigma}{d\Omega}\, K_{j00i} = - \left( \frac{2\,\alpha^2}{t} \right) mM \left[\, S \cdot S' - \frac{S \cdot q \, S' \cdot q}{t}\right]\,.
\end{equation}

 \subsection*{One particle structured}
\noindent
Using the proton electromagnetic form factors as described earlier, and also $S \cdot k = 0$ and $S' \cdot P' = 0$ from the theory of spin polarization,
\begin{eqnarray}
 s\,\frac{d\sigma}{d\Omega}\, K_{j00i}& = 
&\;  -\left( \frac{2\,\alpha^2}{t} \right)mMG_M\left\{F_1\left[S \cdot S' - \frac{S \cdot q\, S' \cdot q}{t}\right]\right. \nonumber \\& 
& \left. \qquad \qquad \qquad \qquad \qquad + \,\frac{F_2}{4M^2} \left[t \, S \cdot S' + 2 \,S \cdot P' \, S' \cdot q \right]\right\} 
\end{eqnarray}
which is consistent with the previous result in the limit $F_1 \rightarrow 1$, $F_2 \rightarrow 0$ and hence $G_M \rightarrow 1$.\\
\\
This represents a relativistic generalization of equation (3) of reference \cite{Horowitz}, where the simplification $s = (M + m)^2 + 2m T \,$
 tending to \( (M + m)^2 \) is made
 in the limit of low laboratory kinetic energy $T$.

\pagebreak

 \subsection*{Both particles structured}
\noindent
Using the electromagnetic form factors $f_1$ and $f_2$ of the second particle as earlier we obtain
\begin{eqnarray}
& &\left( \frac{-t}{2\,\alpha^2} \right)s\,\frac{d\sigma}{d\Omega}\, K_{j00i} = \nonumber\\[2ex]& & = mMg_M^{\,2}G_M\left\{F_1\left[S \cdot S' - \frac{S \cdot q \, S' \cdot q}{t}\right]+ \,\frac{F_2}{4M^2}\left[t \, S \cdot S' + 2 \,S \cdot P' \, S' \cdot q \right]\right\} \nonumber \\[2ex]& &+\,\frac{f_2}{m}\,MG_M^{\,2}g_M\left\{S \cdot S' \left[\,t\,\left(t - 4m^2\right)\right] + \,4 \,m^2S \cdot q \, S' \cdot q + 2\,t \, S \cdot q \, S' \cdot k \right\} \\[2ex]& & + \,\frac{f_2}{m}\,\frac{F_2}{M}\,G_Mg_M k_\tau k'_\alpha  P^\lambda P'^\rho S_\beta S'^\sigma\left[(s-u)\epsilon^{\mu \alpha \beta \tau}\epsilon_{\mu \sigma \rho \lambda}\right. \nonumber \\[2ex] & &\left. \qquad \qquad \qquad \qquad \qquad \qquad \qquad \qquad- \epsilon^{\mu \alpha \beta \tau}\epsilon_{\nu \sigma \rho \lambda}\,\left(k^\nu + k'^\nu\right) \left(P_\mu + P'_\mu\right)\right] \,.\nonumber
\end{eqnarray}
%
%
This expression agrees with the previous formulae in the appropriate limits.  Here the sign convention of the Levi-Civita symbol is defined by $\epsilon_{0123}\,=\,+1$.

 \subsection*{Antiproton proton case}
\noindent
Again the proton and antiproton electromagnetic form factors and masses are the same, and the $s$-channel term is neglected in favour of the $t$-channel contribution in the low momentum transfer region.  Thus
\begin{eqnarray}
& & \left( \frac{-t}{2\,\alpha^2} \right)s\,\frac{d\sigma}{d\Omega}\, K_{j00i}= \nonumber \\[2ex]& & =M^2G_M^{\,3}\left\{F_1\left[S \cdot S' - \frac{S \cdot q \, S' \cdot q}{t}\right] + \frac{F_2}{4M^2}\left[t \, S \cdot S' + 2 \,S \cdot P' \, S' \cdot q \right]\right\} \\[2ex]& &  +\,F_2\,G_M^{\,3}\left\{S \cdot S' \left[\,t\,\left(t - 4M^2\right)\right] + \, 4\,M^2S \cdot q \, S' \cdot q + 2\,t \, S \cdot q \, S' \cdot k \right\}\nonumber \\[2ex]& & + \,\frac{F_2^{\,2}}{M^2}\,G_M^{\,2} k_\tau k'_\alpha  P^\lambda P'^\rho S_\beta S'^\sigma\left\{\epsilon^{\mu \alpha \beta \tau}\left[\epsilon_{\mu \sigma \rho \lambda}(s-u) - \epsilon_{\nu \sigma \rho \lambda}\,\left(k^\nu + k'^\nu\right) \left(P_\mu + P'_\mu\right)\right]\right\} \nonumber
\end{eqnarray}
and again this result is important in the momentum transfer region $|t|<|t_c|$ defined by equation (8),
%
%
%
%
where the electromagnetic interaction dominates the hadronic interaction.

\pagebreak

 \section{Summary}
\noindent
In the proposed PAX facility the transfer of polarization from a polarized proton to repeatedly circulating antiprotons may take place in a region of momentum transfer where the electromagnetic effects compete with the hadronic interaction.  The spin averaged and polarization transfer antiproton proton elastic differential cross sections due to one photon exchange presented here may assist in understanding the spin filtering process that seeks to increasingly polarize antiprotons in a storage ring, in order to measure a number of significant properties of the proton.  In particular, the transversity distribution of the valence quarks in addition to the analytic properties of the time-like electromagnetic form factors would be measured for the first time.

Fully relativistic formulae have been derived for polarization transfer in spin half fermion fermion elastic scattering, due to one photon exchange.  They can be applied to proton electron and antiproton electron scattering, and to near forward (small $t$) antiproton proton scattering where the Coulomb interaction is important.

 \bigskip

 {\small DOB would like to thank the Irish Research Council for Science, Engineering and Technology (IRCSET), for a postgraduate research scholarship.  NHB is grateful to Enterprise Ireland for the award of a grant under the International Collaboration Program to facilitate a visit to INFN at the University of Torino, where discussions with M.~Anselmino and M.~Boglione are acknowledged.}


\bigskip

 \begin{thebibliography}{99}

\bibitem{Rathmann}
F.~Rathmann et al., Phys. Rev. Lett. {\bf 94} (2005) 014801.

\bibitem{Barone:2005pu}
  V.~Barone {\it et al.}  [PAX Collaboration],
  arXiv:hep-ex/0505054.

\bibitem{Milstein:2005bx}
  A.~I.~Milstein and V.~M.~Strakhovenko,
  arXiv:physics/0504183.

\bibitem{Mandelstam}
S.~Mandelstam, Phys. Rev. {\bf 112} (1958) 1344.

\bibitem{Rosenbluth}
M.~N.~Rosenbluth, Phys. Rev. {\bf 79} (1950)  615.

\bibitem{Buttimore:1978ry}
  N.~H.~Buttimore, E.~Gotsman and E.~Leader,
  Phys.\ Rev.\ D {\bf 18} (1978) 694.

\bibitem{Block:1996jd}
  M.~M.~Block,
  Phys.\ Rev.\ D {\bf 54} (1996) 4337.

\bibitem{Lehar:1978}
J.~Bystricky, F.~Lehar and P.~Winternitz, J. Phys. (Paris) {\bf 39} (1978) 1.

\bibitem{Horowitz}
C.~J.~Horowitz and H.~O.~Meyer,  Phys. Rev. Lett. {\bf 72} (1994) 3981.




 \end{thebibliography}
 \end{document}